\documentclass{jetpl}
\twocolumn

\title{Orbital glass and spin glass states of  $^3$He-A in aerogel}

\rtitle{Orbital glass and spin glass states of  $^3$He-A in
aerogel}

\sodtitle{Orbital glass and spin glass states of $^3$He-A in
aerogel}

\author{V.\,V. Dmitriev, D.\,A. Krasnikhin, N. Mulders$^+$,
A.\,A. Senin, G.\,E. Volovik$^{*\Delta}$, A.\,N. Yudin }

\rauthor{V.\,V. Dmitriev, D.\,A. Krasnikhin, N. Mulders, A.\,A.
Senin, G.\,E. Volovik, A.\,N. Yudin}

\sodauthor{V.\,V. Dmitriev, D.\,A. Krasnikhin, N. Mulders, A.\,A.
Senin, G.\,E. Volovik, A.\,N. Yudin}

\address{P.\,L. Kapitza Institute for Physical Problems RAS,
2 Kosygina str., 119334 Moscow, Russia\\~\\
$^+$Department of Physics and Astronomy, University of Delaware,
Newark, Delaware 19716, USA\\~\\ $^*$Low Temperature Laboratory,
Aalto University, P.O. Box 15100, FI-00076 AALTO, Finland \\~\\
$^\Delta$ L.D. Landau Institute for Theoretical Physics, 119334
Moscow, Russia}

\dates{29 April 2010}{*}

\abstract{ Glass states of superfluid A-like phase of $^3$He in
aerogel induced by random orientations of aerogel strands are
investigated theoretically and experimentally. In anisotropic
aerogel with stretching deformation two glass phases are observed.
Both phases represent the anisotropic glass of the orbital
ferromagnetic vector $\hat {\bf l}$ -- the orbital glass (OG). The
phases differ by the spin structure: the spin nematic vector $\hat
{\bf d}$ can be either in the ordered spin nematic (SN) state or
in the disordered spin-glass (SG) state. The first phase (OG-SN)
is formed under conventional cooling from normal $^3$He. The
second phase (OG-SG) is metastable, being obtained by cooling
through the superfluid transition temperature, when large enough
resonant continuous radio-frequency excitation are applied. NMR
signature of different phases allows us to measure the parameter
of the global anisotropy of the orbital glass induced by
deformation. }

\PACS{67.57.Pq, 67.57.Lm, 61.30.Dk, 75.10.Nr}

\begin{document}

\maketitle

\section{Introduction}
Superfluidity of $^3$He in high porosity aerogel \cite{Porto,Halp}
allows to investigate influence of impurities (i.e. aerogel
strands) on superfluidity with nontrivial Cooper pairing. It was
found that like in bulk $^3$He two superfluid phases (called by
analogy A-like and B-like) can exist in $^3$He in aerogel in a
weak magnetic fields \cite{Barker}. The B-like phase is analogous
to bulk $^3$He-B and has the same order parameter
\cite{Barker,Dmit1}. As for the A-like phase then for many years
the situation was not clear. It was proposed that the A-like phase
in aerogel is described by the Larkin-Imry-Ma \cite{Larkin,ImryMa}
(LIM) model \cite{Volovik1}. The random orientations of silicon
strands induces spatially random distribution of $^3$He-A order
parameter. However, properties of nuclear magnetic resonance (NMR)
in various aerogel samples were different and did not correspond
to the properties of the fully randomized bulk A phase. Situation
became much more clear when it was realized that even weak
anisotropy of aerogel can influence the NMR properties and was
found experimentally that in squeezed by $\sim$1\% aerogel the
A-like phase behaves as the bulk $^3$He-A but with the orbital
vector $\hat{\bf l}$ fixed along the deformation axis
\cite{Kunimatsu2007}. This observation was in agreement with LIM
model developed for the case of nonzero global anisotropy
\cite{Volovik3}. It was also found that intrinsic anisotropy in
some samples can be large enough to orient $\hat{\bf l}$ and all
NMR properties of the A-like phase in such samples correspond to
the bulk A phase order parameter oriented along some fixed axis
\cite{Dmitriev2007}.

Here we consider consequences of the theory developed in
\cite{Volovik1,Volovik3} for different values and types of global
anisotropy and report results of detailed NMR investigations of
A-like phase in three aerogel samples with different anisotropy.

\section{Theory}

In superfluid $^3$He, the spin-orbit interaction is small compared
to other characteristic energy scales. That is why the superfluid
phases of $^3$He consist of two nearly independent subsystems of
orbital and spin degrees of freedom. The bulk $^3$He-A is
characterized by nematic ordering in the spin subsystem
\cite{Andreev1984} and by the ferromagnetic ordering in the
orbital subsystem. Its order parameter is the matrix
\begin{equation}
A_{\alpha j}=\Delta \hat d_\alpha (\hat e^1_j + \hat e^2_j) \,.
\label{OrderParameter}
\end{equation}
Here $\hat {\bf d}$ is unit vector describing the nematic spin
order. Orthogonal unit vectors $\hat {\bf e}^1$ and $\hat {\bf
e}^2$ describe orbital ferromagnetism with ferromagnetic moment
along the unit vector $\hat {\bf l}=\hat {\bf e}^1 \times \hat
{\bf e}^2$.

Depending on value and type of anisotropy several possible
structures of the order parameter \eqref{OrderParameter} can be
realized in the A-like phase of $^3$He in aerogel. For isotropic
aerogel, the orbital vector $\hat{\bf l}$ is randomized due to the
quenched local anisotropy provided by random orientations of
aerogel strands \cite{Volovik1}:
\begin{equation}
\left<\hat{\bf l}\right> =0 ~~,~~\left<l_x^2\right> =
\left<l_y^2\right> =\left<l_z^2\right> =1/3~~, \label{LIMstate}
 \end{equation}
This glass state is the realization of the LIM phenomenon in
$^3$He-A. In this state the space average of the order parameter
\eqref{OrderParameter} is zero, $\left<A_{\alpha j}\right>=0$, and
\begin{equation}
\left<A_{\alpha i}A_{\beta j}\right>=0~,~ \left<A_{\alpha
i}A^*_{\beta j}\right>+{\rm c.c.}= \frac{2}{3}\Delta^2
\delta_{ij}\left(\delta_{\alpha\beta}-\hat h_\alpha  \hat
h_\beta\right) \,. \label{RandomOrderParameter}
\end{equation}
Here $\hat {\bf h}$ is unit vector along magnetic field {\bf H}
which keeps $\hat{\bf d}$ in the plane normal to $\hat{\bf h}$,
where $\hat {\bf d}$  is randomized due to spin-orbit interaction
with chaotic orbital momentum $\hat{\bf l}$. We call this
configuration the OG-SG state, since it is the combination of the
orbital glass (OG) and spin-nematic glass (SG). The OG-SG states
produced by random anisotropy of aerogel strands provide the
experimental realization of the random anisotropy glasses
discussed in different systems \cite{Fedorenko2007}, such as
random anisotropy Heisenberg spin glasses in magnets
\cite{Itakura,Dotsenko} and nematic glasses in liquid crystals
\cite{Kats1997,Feldman2004,Feio2008}.

Uniaxial deformation adds more states of superfluid $^3$He-A in
aerogel. Chaotic spatial distribution of the orbital vector
$\hat{\bf l}$ is modified under deformation and becomes
anisotropic. For squeezing or stretching of aerogel we obtain the
orbital glass states with global anisotropy:
\begin{equation}
 \left<\hat{\bf l}\right> =0~~,~~\left<l_z^2\right>=\frac{1+2q}{3}~~,~~\left<l_x^2\right> = \left<l_y^2\right>=\frac{1-q}{3}  \,.
 \label{Nematic}
\end{equation}
Here the axis of deformation is $\hat{\bf z}$ and we introduce
parameter $q$ of global anisotropy, which is positive in squeezed
aerogel, negative in the stretched case and $q=0$ in isotropic
aerogel.

Stretching of aerogel gives the global easy plane anisotropy with
$-0.5<q<0$ and $0<\left<l_z^2\right><1/3$. In the limit of large
stretching, $q$ approaches the value $q=-0.5$, where
$\left<l_z^2\right>=0$, i.e. $\hat{\bf l}$ is kept in
$\bf{\hat{x}-\hat{y}}$ plane and the planar LIM state is formed.
This orbital glass  is described by the random anisotropy $XY$
model.

Squeezing of aerogel gives the global easy axis anisotropy with
$1>q>0$ and $1/3<\left<l_z^2\right><1$. However, in this state $q$
may not reach 1, because for deformations greater than some
critical value \cite{Volovik3} the orbital ferromagnetic (OF)
state should be restored. In the OF state, $\left<\hat{\bf
l}\right> \neq 0$ and is parallel to the deformation axis
$\hat{\bf z}$. At large squeezing, $\left<l_z\right>$ approaches
$+1$ or $-1$. Properties of the OF state correspond to the bulk A
phase but with $\hat{\bf l}$ fixed along the axis of deformation.
Observations of such ferromagnetic state were reported in
\cite{Kunimatsu2007,Dmitriev2007,Sato2008}. For the intermediate
squeezing deformations, the transverse components $\left<
l_x^2\right>$ and $\left< l_y^2\right>$ may be substantial and the
ferromagnetic order, $\left<l_z\right>\neq 0$, is supplemented by
the glass state for $l_x$ and $l_y$ components.

\section{NMR frequency shift}
The frequency shift of transverse NMR from the Larmor value is
given by \cite{Gongadze,VolovikF}:
\begin{eqnarray}
 \Delta \omega/\Delta \omega_{\rm 0}=-\frac{\partial U_D}{\partial  \cos\beta}=
 \nonumber
 \\
 \frac{7 \cos\beta +1}{4}(\hat{\bf l} \times \hat{\bf h})^2-\frac{1+\cos\beta}{2}(\hat{\bf l}\cdot \hat{\bf d}\times \hat{\bf h})^2
 -\cos\beta ~,
\label{FShift}
 \end{eqnarray}
\begin{eqnarray}
U_D=-\frac{1}{2}\sin^2\beta + \frac{1}{4}(1+\cos\beta)^2 (\hat{\bf
l}\cdot \hat{\bf d}\times \hat{\bf h})^2-
\nonumber\\
-(\frac{7}{8}\cos^2\beta +\frac{1}{4}\cos\beta  -\frac{1}{8}) (\hat{\bf l} \times \hat{\bf h})^2
~~,
\label{SO}
 \end{eqnarray}
where $\Delta \omega_{\rm 0}={\Omega_A^2}/({2\omega})$ is maximal
possible value of the shift, $\Omega_A$ is Leggett frequency,
$\beta$ is the tipping angle of magnetization, $U_D$ is the
normalized spin-orbit (dipole-dipole) energy averaged over fast
precession of magnetization. The vector $\hat{\bf d}$ here is the
result of averaging over fast precession, it coincides with the
original spin-nematic vector $\hat{\bf d}$ only for $\beta=0$.

In the LIM state, the dipole energy should be averaged over space.
The LIM characteristic length is expected to be much smaller than
the dipole length, which characterizes the spin-orbit interaction,
so the NMR line should not be broadened due to inhomogeneities of
$\hat{\bf l}$.

We consider the general case when {\bf H} is tilted by an angle
$\mu$ to anisotropy axis $\hat{\bf z}$:
 \begin{equation}
 \hat{\bf h}  =\cos \mu~\hat{\bf z} +\sin \mu~\hat{\bf x} ~.
\label{h}
 \end{equation}
If the global orientation of $ \hat{\bf l}$ is capable to orient
$\hat{\bf d}$ (to minimize $U_D$) then in squeezed aerogel we get:
  \begin{equation}
 \hat{\bf d}  =\sin \mu~\hat{\bf z} -\cos \mu~\hat{\bf x}  ~~,~~q>0  ~,
\label{d>}
 \end{equation}
 and in stretched aerogel:
  \begin{equation}
 \hat{\bf d}  = \hat{\bf y}    ~~,~~q<0    ~.
\label{d<}
 \end{equation}
These states in deformed aerogel combine anisotropic orbital glass
(OG) and the ordered spin nematic (SN) and we denote them as
anisotropic OG-SN states.

If $ \hat{\bf l}$ is not able  to orient $\hat{\bf d}$ then
the chaotic distribution of spin vector $\hat{\bf d}$ is realized:
  \begin{equation}
 \hat{\bf d}  =\cos\Phi_d\left(\sin \mu~\hat{\bf z} -\cos \mu~\hat{\bf x}\right)+\sin\Phi_d ~ \hat{\bf y}
 ~,
\label{drandom}
 \end{equation}
with random $\Phi_d$. This is an anisotropic OG-SG state -- the
anisotropic orbital glass accompanied by spin glass. Note that the
OG and SG  subsystems have different characteristic length scales.

\subsection{
Anisotropic OG-SN  in squeezed aerogel} In this state the spin
nematic is regular, and we should average only over the orbital
glass state. In squeezed aerogel, the averaging over space of
\eqref{d>} gives:
\begin{equation}
\left< (\hat{\bf l}\cdot \hat{\bf d}\times \hat{\bf h})^2\right> =
\left<l_y^2\right>=\frac{1}{2}\left(1- \left<l_z^2\right>\right)~,
\label{average1>}
 \end{equation}
\begin{equation}
\left< (\hat{\bf l}\times \hat{\bf h})^2\right> =
1-\left<l_z^2\right>+\frac{1}{2}\sin^2\mu \left(3\left<l_z^2\right>-1\right)~.
\label{average2>}
 \end{equation}
Then the NMR frequency shift is
 \begin{equation}
\frac{\Delta \omega}{\Delta \omega_{\rm 0}}
=q\left(-\cos\beta+\sin^2\mu \frac{7\cos\beta +1}{4}\right)~~.
\label{generalH}
 \end{equation}
Equation \eqref{generalH} with $q\approx 1$ is applicable also for
the orbital ferromagnetic (OF). Therefore NMR can not distinguish
between OF and OG-SN states in squeezed aerogel with $q=1$. The
latter state may be considered as multi-domain OF state or the
Ising orbital glass.

\subsection{Anisotropic OG-SN  in  stretched aerogel} For stretched aerogel,
from \eqref{d<} we obtain:
\begin{equation}
\left< (\hat{\bf l}\cdot \hat{\bf d}\times \hat{\bf h})^2\right> =\frac{1}{2}\left(1- \left<l_z^2\right>\right)+\frac{1}{2}\sin^2\mu \left(3\left<l_z^2\right>-1\right)~,
\label{average1<}
 \end{equation}
and the corresponding NMR frequency shift is
 \begin{equation}
\frac{\Delta \omega}{\Delta \omega_{\rm 0}}
=q\left(-\cos\beta+\sin^2\mu \frac{5\cos\beta -1}{4}\right)~~.
\label{oriented_q<0}
 \end{equation}

\subsection{Anisotropic orbital glass + spin glass}
In the OG-SG state the spin-nematic vector $\hat{\bf d}$ in
\eqref{drandom} is random, and we obtain:
\begin{equation}
\left< (\hat{\bf l}\cdot \hat{\bf d}\times \hat{\bf h})^2\right>  =\frac{1}{2}
\left< (\hat{\bf l}\times \hat{\bf h})^2\right> ~,
\label{AverageRandomD}
 \end{equation}
with $\left< (\hat{\bf l}\times \hat{\bf h})^2\right>$ from
\eqref{average2>}. Then $\Delta\omega$ is given by:
\begin{eqnarray}
\frac{\Delta \omega}{\Delta \omega_{\rm 0}} =q\cos\beta\left(
\frac{3}{2}\sin^2\mu-1\right)~, \label{AverageRandomD}
 \end{eqnarray}
which is valid for both squeezed and stretched aerogel.

\begin{figure}[h]
\begin{center}
\includegraphics[scale=0.35]{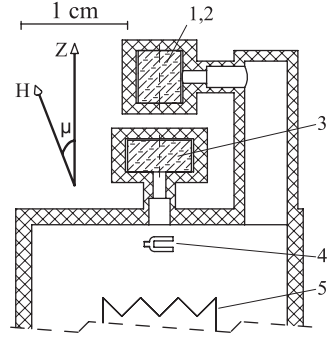}
\end{center}
\caption{Fig.1. Sketch of experimental chamber. 1,2 - samples 1 or
2; 3 - sample~3; 4 - quartz tuning fork; 5 - heater.} \label{f1}
\end{figure}

\section{Conditions of experiments}
Three different aerogel samples with porosity of 98.2\% and with
different types and values of anisotropy were used. All the
samples had cylindrical form with axes oriented along $\hat{\bf
z}$. Diameter and height of the samples (in mm) were the
following: 4 and 4 (sample~1), 3.8 and 5 (sample~2), 6 and 3
(sample~3). The experimental chamber (Fig.1) was made from epoxy
resin "Stycast-1266". The upper cell (1 and 2 on Fig.1) was used
first for sample~1 and then for sample~2. Samples 1 and 2 had gaps
$\sim$0.1 mm from side walls of the cell but were squeezed at room
temperature along $\hat{\bf z}$-axis by 8\% and 3\% respectively,
i.e. after cooling their deformation should be 9\% and 4\% due to
additional thermal shrinkage of epoxy walls by $\sim$1\%. Sample~3
was placed freely in the cell with the gaps $\sim$0.1 mm from side
and top walls to avoid such additional deformation. Its residual
anisotropy was measured at room temperature by light birefringence
\cite{Mulders} and was less than 0.1\%.

In order to avoid paramagnetic signal from solid $^3$He, aerogel
samples were preplated by $\sim$2.5 monolayers of $^4$He. Each
sample was surrounded by transverse NMR coil (not shown in Fig.1)
with the axis along $\hat{\bf x}$. We were able to rotate $\bf H$
by any angle in $\hat{\bf y}-\hat{\bf z}$ plane and also by
$\pm15^\circ$ in $\hat{\bf x}-\hat{\bf y}$ plane. Experiments were
carried out in magnetic fields from 95 to 424 Oe (corresponding
NMR frequencies are from 310.5 KHz to 1.38 MHz) and at pressures
26.0 (sample~1) and 27.2 bar (samples~2 and 3). Necessary
temperature was obtained by nuclear demagnetization cryostat and
was measured by quarz tuning fork, calibrated by Leggett frequency
measurements in bulk $^3$He-B.

\begin{figure}[h]
\begin{center}
\includegraphics[scale=0.9]{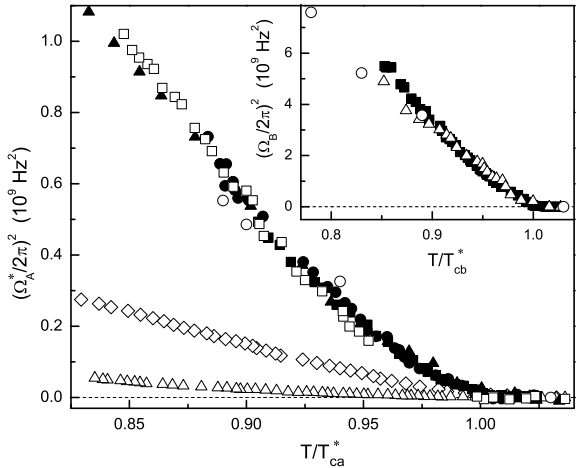}
\end{center}
\caption{Fig.2. Temperature dependence of the "effective" Leggett
frequency in the A-like phase and $\Omega_B$ in B-like phase
(insert) in different aerogel samples rescaled to 26 bar pressure.
({\scriptsize$\square$}) - sample~1 (9\% squeezing, 26.0 bar);
($\blacktriangle$) - squeezed by 4\% aerogel (29.3 bar)
\cite{Kunimatsu2007}; ({\Large$\bullet$}) - intrinsically
anisotropic aerogel (28.6 bar) \cite{Dmitriev2008};
({\scriptsize$\blacksquare$}) - intrinsically anisotropic aerogel
(26.0 bar) \cite{Dmitriev2007}; ({\Large{$\circ$}}) - aerogel
squeezed radially by 20\% (25.0 bar) \cite{Elbs2008};
({\Large$\diamond$}) - sample~2 (27.2 bar); ({\small$\triangle$})
- sample~3 (27.2 bar).} \label{f2}
\end{figure}

\section{Continuous wave NMR experiments}
For sample~1 and for ${\bf H}\parallel\hat{\bf z}$, the continuous
wave (CW) NMR shift in the A-like phase was negative as it is
expected from \eqref{generalH} for $\mu$=0. For 9\% squeezing we
should get the orbital ferromagnetic phase with $q\approx 1$,
since from \cite{Kunimatsu2007} it follows that the critical value
of anisotropy is less than 1\%. Therefore from the value of the
shift we can extract the Leggett frequency for this sample.
Temperature of superfluid transition $T_{ca}$ of $^3$He in aerogel
is not well defined. In particular, on warming NMR shift in the
B-like phase disappears at temperature by $\sim 0.02\,T_{ca}$
higher than in A-like phase \cite{Dmitriev2007}. Therefore, to
compare our data with results of previous experiments
\cite{Kunimatsu2007,Dmitriev2007,Dmitriev2008,Elbs2008} we defined
$T_{ca}^{*}$ as the temperature when the temperature dependence of
the NMR frequency shift in the A-like phase is extrapolated to
zero. Then all the data were recalculated to the same pressure
(26.0 bar) using pressure dependence of the Leggett frequency in
bulk $^3$He-A \cite{Hels}. Correction coefficients for the shift
were 0.854 (for 29.3 bar), 0.881 (28.6 bar), 0.949 (27.2 bar) and
1.066 (for 25 bar). Fig.2 shows the resulting "effective" Leggett
frequency $(\Omega_A^*)^2=2\omega|\Delta\omega|$ obtained in these
CW NMR experiments ($\beta=0$). It is seen that all the data
(except for samples~2 and 3) collapse into the same curve. This
suggests that the orbital ferromagnetic state with $q\approx 1$
was obtained in these experiments. Correspondingly the obtained
$\Omega_A^*(T)$ is in fact the true Leggett frequency
$\Omega_A(T)$. The values of $T_{ca}^{*}$ for our samples were
found to be 0.74\,$T_c$ (sample~1, 26.0 bar), 0.722\,$T_c$
(sample~2, 27.2 bar), 0.77\,$T_c$ (sample~3, 27.2 bar), where
$T_c$ is the superfluid transition temperature of the bulk $^3$He.

Measurements of the Leggett frequency in the B-like phase
($\Omega_B$) were also done with sample~3 using a prominent kink
at the low field region of the CW NMR line. The results were
compared with the B-like phase data published in
\cite{Dmitriev2007,Elbs2008}. Here we also have recalculated the
data to pressure of 26 bar using pressure dependence of the
Leggett frequency in B-like phase measured in \cite{LeggB}
(correction coefficients were 1.026 for 25 bar and 0.983 for 27.2
bar). It is seen that these data also collapse to a single curve
(insert in Fig.2) if we define $T_{cb}^*\approx T_{ca}$ as the
temperature when the frequency shift in the B-like phase is
extrapolated to zero.

\begin{figure}[h]
\begin{center}
\includegraphics[scale=0.9]{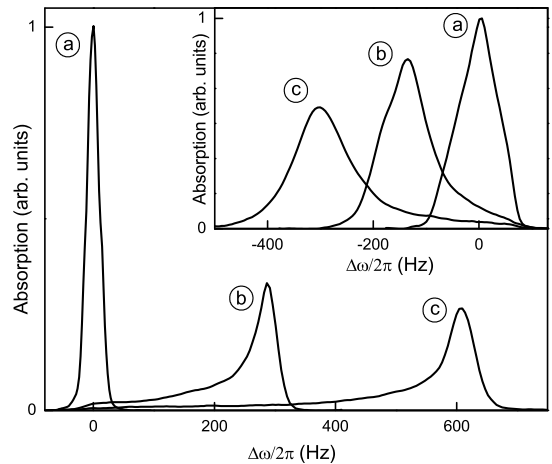}
\end{center}
\caption{Fig.3. NMR absorption lines in sample~2 in the OG-SN
state (orbital glass combined with regular spin nematic) for ${\bf
H}\parallel{\bf{\hat z}}$. NMR frequency is 328.5 kHz. a --
$T>T_{ca}^{*}$;  b -- $T=0.88\,T_{ca}^{*}$; c --
$T=0.77\,T_{ca}^{*}$. Insert: NMR lines in the  OG-SG state
(orbital glass accompanied by spin glass) for $\mu=90^\circ$ at
corresponding conditions. } \label{f1}
\end{figure}

In contrast to sample~1, frequency shifts in samples 2 and 3 for
${\bf H}\parallel\hat{\bf z}$ were much smaller and positive (see
Fig.2 and the mainframe Fig.3). It can be explained if we suggest
that in both cases we have stretched aerogel with the anisotropic
OG-SN state, i.e. the orbital glass combined with regular spin
nematic. Rotation of {\bf H} in $\hat{\bf x}-\hat{\bf y}$ plane
did not change NMR properties, i.e. the observed state is nearly
isotropic in this plane.

We assume that in both samples the Leggett frequency equals to the
value which follows from the most of the data in Fig.2. Then from
the measured values of $\Delta\omega$ we obtain that
$q\approx-0.25$ for sample~2 and $q\approx-0.05$ for sample~3,
i.e. the sample~3 is close to isotropic aerogel. The fact that
sample~2 was squeezed by $\sim$4\% but behaves like stretched
sample can be explained by large initial intrinsic stretching:
this sample was grown in quartz tube with i.d. of 4 mm, but for
unknown reason it was shrunk in diameter by about 5-6\% after
preparation.

\begin{figure}[h]
\begin{center}
\includegraphics[scale=0.9]{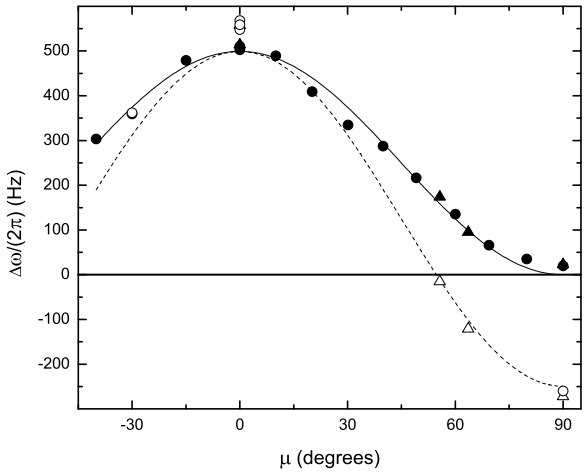}
\end{center}
\caption{Fig.4. Dependence of CW NMR frequency shift in the A-like
phase (sample~2) on the angle $\mu$ between sample axis and $\bf
H$. Filled symbols were obtained after cooling through $T_{ca}$
with low RF field and open symbols -- with high RF excitation. NMR
frequency is 328.5 KHz, but data shown by triangles were obtained
at 341.5 KHz and were rescaled to 328.5 KHz. Solid line is the
best fit of filled symbols by \eqref{oriented cw}, while the
dashed line is the theoretical dependence \eqref{AverageRandomD}.
$T=0.81\,T_{ca}^*$. } \label{f3} \end{figure}

To confirm that in samples 2 and 3 we have stretched aerogel with
the orbital glass state, we have measured CW NMR shifts for
different angles $\mu$ between $\hat{\bf z}$ and $\bf{H}$. For the
anisotropic OG-SN state in stretched aerogel (see
\eqref{oriented_q<0}) and for CW NMR ($\cos\beta$=1) we get
\begin{equation}
{\Delta \omega}=A\cos^2\mu\~~, \label{oriented cw}
\end{equation}
where $A=-q\Delta\omega_{\rm 0}$. Note that for stretched aerogel
one has $q<$0, i.e. $\Delta\omega$ is always positive. The
obtained dependence of $\Delta\omega$ versus $\mu$ for sample~2 is
shown in Fig.4 by filled symbols. Solid line is the best fit by
theoretical dependence \eqref{oriented cw} with $A$=499 Hz. It
gives us the same $q=-0.25$ for this sample as obtained from
Fig.2.

The orbital glass with the limiting value $q=-0.5$  was first
introduced in \cite{Elbs2008}. This  planar  LIM state  of
$^3$He-A was suggested to explain the NMR experiments in aerogel
obtained by 20\% radial squeezing of cylindrical sample, which
would correspond to a large stretching deformation needed to form
the  $XY$ orbital glass. The identification was based on the
comparison of the measured NMR frequency shifts in  A- and B-like
phases which was found to be about twice less than in bulk $^3$He.
However, according to Fig.2, the value of $\Omega_A^*$ obtained in
\cite{Elbs2008} more likely corresponds to the orbital
ferromagnetic state, rather than to the planar LIM state. In
principle, a small inhomogeneity in radial deformation can destroy
the planar LIM state in the most parts of the sample and restore
the ferromagnetic order with the anisotropy axis laying in the
$\hat{\bf x}-\hat{\bf y}$ plane.  As for the ratio of
$\Omega_A^2/\Omega_B^2$ then it seems that in $^3$He in aerogel it
is less than in bulk superfluid (see Fig.2).

The data described above were obtained if low level of
radio-frequency (RF) excitation was used to monitor the NMR line
on cooling through $T_{ca}$. Earlier it was found
\cite{Dmitriev2006,Dmitriev2008} that the A-like phase may exist
in two states. One state is obtained if on cooling through
$T_{ca}$ the low level of RF (or no RF) is used. Another one is
created by violent perturbation of the spin system during
cooling: when cooling  through $T_{ca}$ is accompanied either by a
very high CW RF excitation (which saturates NMR signal in normal
phase) or by large tipping pulses. We suggest that in this case
the OG-SG state (the orbital glass combined with the spin glass)
can be created. The point is that saturation of the NMR line just
below $T_{ca}$ (where the dipole energy is small) results in fast
rotation of $\hat{\bf d}$. Further cooling can "freeze" the
chaotic distribution of $\hat{\bf d}$ in the plane normal to $\bf
H$.

To check this possibility we carried out "high RF cooling through
$T_{ca}$" experiments with samples 2 and 3. The CW RF field was
kept $\sim$0.015 Oe which is about 50 times larger than usual, so
that the NMR line in normal phase was fully saturated. After
cooling below $\sim 0.9\,T_{ca}$, the RF level was reduced to its
normal value. NMR lines obtained in such experiment for
$\mu=90^\circ$ are shown in the insert to Fig.3. It was found that
the state which appears after such cooling well corresponds to the
OG-SG state in stretched aerogel. By open symbols in Fig.4 are
shown results obtained in the OG-SG state in sample~2 in CW NMR.
This can be compared with the filled symbols obtained for the
spin-nematic state (OG-SN) in the same sample at the same
conditions. Dashed curve is the theoretical dependence expected
for spin glass, $\Delta\omega=-A\left(\frac{3}{2}
\sin^2\mu-1\right)$(see Eq. \eqref{AverageRandomD} for
$\cos\beta$=1). There are no fitting parameters here because the
prefactor $A$ is already found to be 499 Hz at this temperature
from measurements on the OG-SN state. There is a small discrepancy
between theory and experiment for $\mu=0$. This may occur if the
characteristic length of the spin glass is comparable with the
dipole length. Then an additional shift connected with the
gradient energy of the $\hat{\bf d}$-field distribution appears.

In conclusion to this section we should mention that similar
dependencies as shown in Fig.4 were obtained in sample~3 but
values of the shifts were smaller in accordance to the smaller
value of $q$ in this sample. The width of the NMR line in sample~3
was of the order of the shift, i.e. stretching of this sample
seems to be inhomogeneous. It means that the obtained value of $q$
is in fact the value averaged over the sample volume.

\begin{figure}[h]
\begin{center}
\includegraphics[scale=0.9]{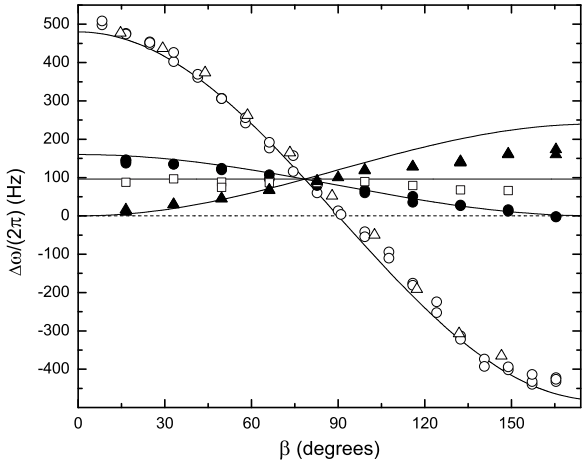}
\end{center}
\caption{Fig.5. Frequency of FIDS in the A-like phase (sample~2)
versus $\beta$ for different $\mu$ in case when low RF excitation
was used while cooling through $T_{ca}$. ({\Large{$\circ$}}) and
({\small$\triangle$}) - $\mu=0$; ({\Large$\bullet$}) -
$\mu=54.7^\circ$; ({\scriptsize$\square$}) -
$\mu=\mu_c=63.4^\circ$; ($\blacktriangle$) - $\mu=90^\circ$. Solid
lines are theoretical curves for corresponding $\mu$ with no
fitting parameters. NMR frequency is 341.5 KHz (except
{\small$\triangle$}, which were obtaned at 664 KHz but rescaled to
341.5 KHz). $T=0.81\,T_{ca}^*$.} \label{f4}
\end{figure}

\begin{figure}[h]
\begin{center}
\includegraphics[scale=0.9]{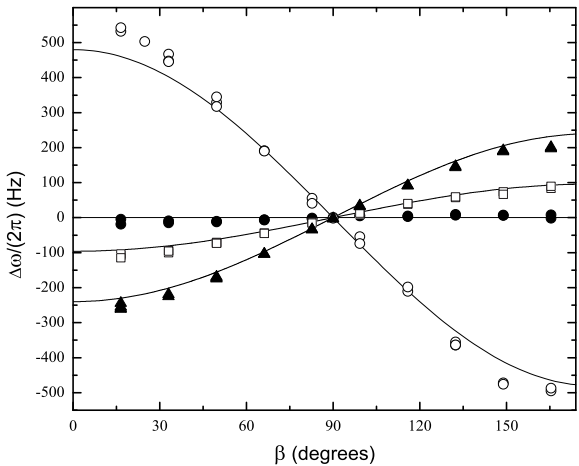}
\end{center}
\caption{Fig.6. Frequency of FIDS in the A-like phase (sample~2)
versus $\beta$ for different $\mu$ in case when very high RF
excitation was used while cooling through $T_{ca}$.
({\Large{$\circ$}}) - $\mu=0$; ({\Large$\bullet$}) -
$\mu=\mu_c=54.7^\circ$; ({\scriptsize$\square$}) -
$\mu=63.4^\circ$; ($\blacktriangle$) - $\mu=90^\circ$. Solid lines
are theoretical curves for corresponding $\mu$ with no fitting
parameters. NMR frequency is 341.5 KHz, $T=0.81\,T_{ca}^*$.}
\label{f5} \end{figure}

\section{Pulsed NMR experiments}
Samples 2 and 3 were also investigated by pulsed NMR. Below we
present the results of experiments with sample~2 which were more
systematic, but we should mention that results for sample~3 were
similar.

From \eqref{oriented_q<0} and \eqref{AverageRandomD} we can obtain
dependencies for $\Delta\omega$ on $\beta$ for different
orientations of $\bf H$. The remarkable feature of
\eqref{oriented_q<0} and \eqref{AverageRandomD} is that we can
introduce the critical angle $\mu_c$. If $\mu < \mu_c$ then the
frequency of free induction decay signal (FIDS) should decrease
with $\beta$ in range $0<\beta<180^\circ$, while for $\mu > \mu_c$
the frequency should grow with increase of $\beta$. For
$\mu=\mu_c$ no dependence of FIDS frequency on $\beta$ is
expected. The value of $\mu_c$ is different for the OG-SN and
OG-SG states (the states with oriented and disordered spin nematic
vector respectrively). In the  OG-SN state, from
\eqref{oriented_q<0} we get that $\sin^2\mu_c=4/5$ (i.e. $\mu_c
\approx 63.4^\circ$),  while for the OG-SG state from
\eqref{AverageRandomD} we get $\sin^2\mu_c=2/3$ and $\mu_c \approx
54.7^\circ$.

After application of single RF tipping pulse the FIDS was
amplified and then recoded by digital oscilloscope. In order to
determine FIDS frequency for a given $\beta$, the obtained time
dependence of the frequency was extrapolated to the initial
moment. The RF pulse was short (15 periods at frequency $\sim$300
KHz), so its spectral width was large enough. The measured
dependencies $\Delta\omega(\beta)$ for different $\mu$ are shown
in Fig.5 for OG-SN state and in Fig.6 for OG-SG state. In both
figures the theoretical functions $\Delta\omega(\beta)$ for
corresponding values of $\mu$ are drawn by solid lines. No fitting
parameters were used for these lines. The parameter
$A=-q\Delta\omega_{\rm 0}$=480 Hz was obtained from its value
found in CW measurements ($A$=499 Hz) after rescaling to the
frequency  341.5 KHz.

It is seen that deviations from theoretical curves in Fig.5 are
small. These deviations are more prominent at large tipping angles
that can be due to inhomogeneity of RF field which was estimated
to be about 15\%. Deviations in Fig.6 for $\mu$=0 and small
$\beta$ are clearly due to the same reason as in CW NMR (see
Fig.4).

\section{Discussion}
Two states of $^3$He-A in aerogel have been discussed here, OG-SN
and OG-SG. Both represent orbital glass (OG) -- the anisotropic
glass state of the orbital vector $\hat {\bf l}$. They differ by
the structure of the spin subsystem, which is either in the spin
nematic (SN) phase or in the disordered spin-glass (SG) state. The
OG-SG state was obtained by cooling through $T_c$ when a large
resonant CW RF excitation was applied, while the OG-SN phase is
formed under conventional cooling through $T_c$. We observe both
these states in each of two stretched aerogel samples: with small
and large deformations.

Leggett frequency in A-like phase of $^3$He in 98.2\% aerogel was
measured using squeezed sample. The results are in a good
agreement with previous data. This suggests that the Leggett
frequency is nearly the same for all samples and to estimate the
deformation parameter $q$ for stretched aerogel samples, which
characterizes the uniaxial global anisotropy of orbital glass.

Most of the previous NMR results in A-like phase can be explained
in frames of the Larkin-Imry-Ma random anisotropy model. In
particular, "soliton-like" spin state observed in
\cite{Dmitriev2008} corresponds to the orbital ferromagnetic state
accompanied by the spin glass state (OF-SG). It is also now clear
that aerogel samples can be intrinsically stretched. It is
possible that in some samples we have both stretched and squeezed
regions. Such situation is more probable in weakly anisotropic
samples. It is also possible that some samples have bi-axial
global anisotropy with $\left<l_x^2\right> \ne \left<l_y^2\right>
\ne \left<l_z^2\right>$. Then we can obtain other than
\eqref{generalH}, \eqref{oriented_q<0} or \eqref{AverageRandomD}
dependencies for $\Delta\omega$ on $\beta$. For example, in the
bi-axial OG-SN state with $\left<l_z^2\right>\approx 1/3$, and
$\left<l_x^2\right> \ne \left<l_y^2\right>$ for $\mu$=0 one gets
$\Delta\omega \propto (1+\cos\beta)$. Such dependence was observed
in weakly anisotropic aerogel samples in
\cite{Ishikawa,Dmitriev2006,Dmitriev2007}.

We conclude that the LIM model for the orbital glass state of the
A-like phase is in a good agreement with the most of experimental
data. Correspondingly the model of "robust" superfluid phase for
the A-like phase suggested in \cite{Fomin} is not realized in
practice at least for values of anisotropy $\geq$0.1\%. However,
many questions concerning the superfluid $^3$He in aerogel still
remain. For example, it is not clear why the NMR frequency shift
in the B-like phase disappears at temperature higher than in the
A-like phase, and why the ratio of Leggett frequencies in the
A-like and B-like phases at the same conditions is less than the
same ratio for bulk A and B phases. It is also not known how $q$
is related to the magnitude of deformation. The transition from
the orbital glass to the orbital ferromagnetic state was not yet
observed and the value of critical anisotropy has not been
measured. This is the subject of further investigations.

This work was supported in part by RFBR grants (09-02-01185 and
09-02-12131ofi) and by the Academy of Finland, Centers of
excellence program 2006-2011. It is a pleasure to thank E. Kats
for helpful comments.

\end{document}